\makeatletter\def\graphicscache@inhibit{true}\makeatother
\documentclass[runningheads]{llncs}
\usepackage[T1]{fontenc}
\usepackage{graphicx}

\usepackage[backend=biber,style=apa,url=true,natbib=true]{biblatex}
\usepackage{xurl}  %

\usepackage{tikz}
\usetikzlibrary{calc}
\usepackage{subfigure}

\usepackage{pgf-umlcd}

\usepackage{siunitx}

\usepackage{todonotes}

\usepackage{tabularx}
\usepackage{booktabs}

\usepackage[noabbrev,nameinlink]{cleveref}

\definecolor{darkred}{RGB}{166, 0, 0}
\definecolor{darkgreen}{RGB}{19, 166, 0}

\newcommand{\millisec}[1]{\SI{#1}{\milli\second}}

\begin{document}
\title{A Service Architecture for Dataspaces}
\author{Benedikt T. Arnold\inst{1,2}\and%
    Christoph Lange\inst{1,2}\and
    Christina Gillmann\inst{1,2}\and
    Stefan Decker\inst{1,2}
}
\authorrunning{B.T. Arnold et al.}
\institute{Fraunhofer Institute for Applied Information Technology FIT, Sankt Augustin, Germany\and
RWTH Aachen University, Aachen, Germany\\
\email{benedikt.arnold@fit.fraunhofer.de}\\
}
\maketitle              %
\begin{abstract}
Dataspaces are designed to support sovereign, trusted and decentralized data exchange between participants forming an ecosystem. They are standardized by initiatives such as the International Data Spaces Association or Gaia-X and have gained adoption in several domains such as mobility, manufacturing, tourism or culture. In dataspaces, participants use connectors to communicate peer-to-peer. The Eclipse Dataspace Components (EDC) Connector is a broadly adopted, open-source implementation that adheres to the standards and is supported by a large community. As dataspaces in general, it focuses on the exchange of data assets with associated usage policies and does not support services. In practice, however, there is demand for dataspace-based services and conceptual arguments support their inclusion in dataspaces. In this paper, we propose an abstraction layer for providing generic services within dataspaces. Adopters can use this layer to easily develop own services, seamlessly integrated with the existing dataspace technology. Besides, we present an initial implementation of this service architecture for the EDC Connector and demonstrate its practical applicability.

\keywords{Dataspaces \and Services \and Decentralized Ecosystems}
\end{abstract}
\section{Introduction}
\label{sec:intro}
Dataspaces have first been introduced by~\citet{franklin2005databases} in 2005, with different scientific definitions over time, collected by~\citet{theissen2023semantics}. Ten years ago, with the initial \emph{Industrial Data Space} research project, kicked-off in 2015~\citep{otto2019designing}, dataspace standardization and adoption has started. Originating from this project, the \emph{International Data Spaces Association} (\emph{IDSA} or \emph{IDS}) has been established through generalization, internationalization and renaming. Since then, many verticalizations following the IDS~\citep{otto2023ram4} and \emph{Gaia-X}~\citep{gaiax2025architecture} specifications have emerged. The core component of interaction between organizations participating  in a dataspace is the \emph{connector}. A widely adopted open-source connector implementation is the \emph{Eclipse Dataspace Components Connector} (\emph{EDC Connector}, short: \emph{EDC}) which is under governance of the Eclipse Foundation, is actively developed, and has extensibility as a core concept.

The dataspace connector technology is mostly focused on the exchange of data assets, each associated with access and usage policies. After establishing a contract for one of these assets, the two related participating organizations can transfer the actual data, which might be either static or dynamic data, served via, e.g., a \emph{REST API}. With the EDC, transfers can be parameterized by the consumer and, recently, a back-channel from the data consumer to the provider has been added to the connector's architecture. Still, the existing tooling allows for services only by wrapping them into data asset-like transfers, requiring significant technical effort to provide a service.

However, adding (data) services to dataspaces offers significant benefits: First, this removes the need to share ``raw data'' and allows for returning only some of the data given inputs. In a similar way, services also allow for completely keeping data confidential and only using it to produce a result, like parameter sets for a neural network: instead of sharing the model, the model owner can return model predictions as a result for incoming input data from the service consumer -- be it image classification or retrieval-based \emph{Large Language Model (LLM)} setups. Besides, services can help addressing the intentionally omitted yet often required data integration in dataspaces by means of data format translation services. While these services are also possible to implement without dataspaces, implementing them in the context of dataspaces yields the associated benefits such as technically established trust, single interaction endpoints and transaction logging.

Beyond the conceptual benefits of such a service architecture, there is significant demand, as experienced in research projects such as Germany's Culture Dataspace project or the DIANA-T project\footnote{``Digitale Leistungen, Datenintegration und Datenautonomie für eine nachhaltigere Tourismusbranche'', eng.: ``Digital Services, Data Integration and Data Sovereignty for a more sustainable Tourism Industry'', \url{https://diana-t.de/}}, and many use cases cannot be realized with dataspaces in their current form without services.

To facilitate straightforward service provision via dataspaces, we introduce an abstraction layer, giving service providers an easy-to-use interface against which they can implement their services. We propose to consider services as first-class citizen dataspace assets, associated with usage policies. The service layer abstracts away the difficulties of decentralized, asynchronous communication between the service consumer and provider as well as establishing a contract for using the service.

Our main contribution is a \textbf{novel abstraction layer for dataspace connectors that allows for providing and consuming arbitrary data services} $f: \left(x_{1}, \ldots, x_{n}\right) \mapsto y$ that map $n$ input arguments $x_{i}$ to an output value $y$ and a prototypic implementation for the EDC Connector.

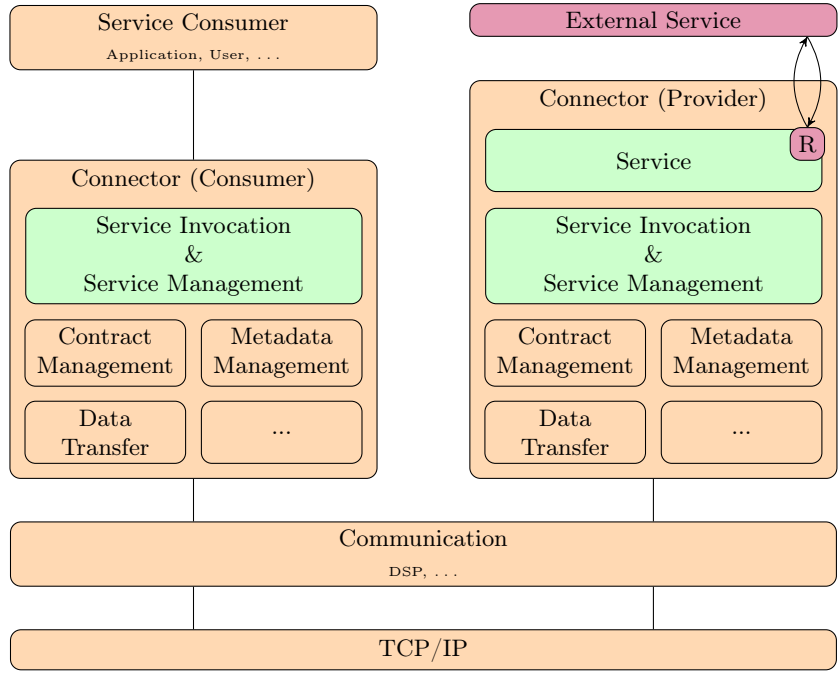
\begin{figure}[ht]
    \centering
    \tikzstyle{component} = [rectangle, draw=black, rounded corners, text centered, fill=orange!30, align=center]
    \tikzstyle{innercomp} = [component,minimum width=.175\textwidth, minimum height={2em+5px}]
    \tikzstyle{servicelayer} = [innercomp, fill=green!20]
    \tikzstyle{bentarrow} = [bend left=30]
    \begin{tikzpicture}[node distance=1cm,>=stealth']
        \node (tcpip) [component, minimum width=.9\textwidth] {TCP/IP};
        \node (dsp) [component, minimum width=.9\textwidth] at ([xshift=0cm,yshift=1cm]tcpip.north) {Communication\\ {\tiny DSP, $\ldots$}};

        \node (consumercon1) [component, above of=dsp, minimum height=120px, minimum width=.4\textwidth, xshift=-.05\textwidth, anchor=south east,label={[shift={(0ex,-4ex)}]north:Connector (Consumer)}] {};
        \node (consumercon2) [component, above of=dsp, minimum height=150px, minimum width=.4\textwidth, xshift=.05\textwidth,  anchor=south west,label={[shift={(0ex,-4ex)}]north:Connector (Provider)}] {};

        \node (datatrans) [innercomp, anchor=south west] at ([xshift=.2cm,yshift=.2cm]consumercon1.south west) {Data\\Transfer};
        \node (etc) [innercomp, anchor=south east]       at ([xshift=-.2cm,yshift=.2cm]consumercon1.south east) {...};

        \node (contract) [innercomp, anchor=south west] at ([xshift=.2cm,yshift={.2cm+2em+5px+.2cm}]consumercon1.south west) {Contract\\Management};
        \node (meta) [innercomp, anchor=south east]     at ([xshift=-.2cm,yshift={.2cm+2em+5px+.2cm}]consumercon1.south east) {Metadata\\Management};

        \node (service) [servicelayer, anchor=south east,minimum width={.35\textwidth+.2cm}]     at ([xshift=-.2cm,yshift={.2cm+2em+5px+.2cm+2em+5px+.2cm+2px}]consumercon1.south east) {Service Invocation\\\&\\Service Management};%

        \node (datatrans2) [innercomp, anchor=south west] at ([xshift=.2cm,yshift=.2cm]consumercon2.south west) {Data\\Transfer};
        \node (etc2) [innercomp, anchor=south east]       at ([xshift=-.2cm,yshift=.2cm]consumercon2.south east) {...};

        \node (contract2) [innercomp, anchor=south west] at ([xshift=.2cm,yshift={.2cm+2em+5px+.2cm}]consumercon2.south west) {Contract\\Management};
        \node (meta2) [innercomp, anchor=south east]     at ([xshift=-.2cm,yshift={.2cm+2em+5px+.2cm}]consumercon2.south east) {Metadata\\Management};

        \node (service2) [servicelayer, anchor=south east,minimum width={.35\textwidth+.2cm}]     at ([xshift=-.2cm,yshift={.2cm+2em+5px+.2cm+2em+5px+.2cm+2px}]consumercon2.south east) {Service Invocation\\\&\\Service Management};%
        \node (service2s) [servicelayer, anchor=south east,minimum width={.35\textwidth+.2cm}]     at ([xshift=-.2cm,yshift={.2cm+2em+5px+.2cm+2em+5px+.2cm+3em+5px+.2cm+2px+2px+2px}]consumercon2.south east) {Service};%

        \node (r) [rectangle,draw=black,rounded corners,text centered, fill=purple!40,align=center] at ([xshift=-.2cm,yshift=-.2cm]service2s.north east) {R};
        \node (ext) [rectangle,draw=black,rounded corners,text centered, fill=purple!40,align=center, minimum width=.4\textwidth] at ([xshift=0cm,yshift=.8cm]consumercon2.north) {External Service};

        \node (cons) [component, minimum width=.4\textwidth] at ([xshift=0cm,yshift={1.8cm-5px}]consumercon1.north) {Service Consumer\\ {\tiny Application, User, $\ldots$}};

        \draw[->] (r.north) to[bentarrow] (r.north |- ext.south);
        \draw[->] (r.north |- ext.south) to[bentarrow] (r.north);

        \draw (consumercon1.south) -- ( consumercon1.south |- dsp.north);
        \draw (consumercon1.south |- dsp.south) -- (consumercon1.south |- tcpip.north);
        \draw (consumercon2.south) -- ( consumercon2.south |- dsp.north);
        \draw (consumercon2.south |- dsp.south) -- (consumercon2.south |- tcpip.north);
        \draw (cons.south) -- (cons.south |- consumercon1.north);
    \end{tikzpicture}
    \caption{Our service abstraction layer in the larger architecture: As part of the connector, our service abstraction layer provides the consumer with an interface to call services and the provider with an interface to offer services, optionally external ones. It abstracts the communication between consumer and provider, which happens via the inter-connector Dataspace Protocol (DSP, \citep{idsa2024dsp}) and on a lower level TCP/IP while enhancing the communication with authentication and trust. The red ``R'' indicates a wrapper service that invokes a remote, external service.}
    \label{fig:layer}
\end{figure}

As depicted in \cref{fig:layer}, our service abstraction layer is part of the connector and, as described before, encapsulates the communication between service consumer and provider.

The rest of this paper is structured as follows: in \cref{sec:relatedwork}, we characterize the relation to the IDS and Gaia-X specifications as well as the relation to the the EDC and give an overview of related approaches. \Cref{sec:architecture} describes our service abstraction layer architecture in detail. The proof-of-concept implementation for the EDC is presented in \cref{sec:impl} and evaluated in \cref{sec:evaluation}. This section is followed by a critical discussion of both the evaluation results and the design decisions, pointing to further required research. \Cref{sec:conclusion} concludes our work and gives an outlook to future research.

\section{Background and Related Work}
\label{sec:relatedwork}
\subsection{Relation to the International Data Spaces}
The International Data Spaces Association (IDSA) is a dataspace initiative and pre-standardization body, emerged from Fraunhofer's Industrial Data Space research projects. Dataspaces are promoted as a concept for establishing \emph{data sovereignty} by the IDSA. A key publication is their \emph{IDS Reference Architecture Model} (\emph{RAM}), highlighting technical as well ecosystem- and governance-related aspects, see chapter 1 of its current version 4.2~\citep{otto2023ram4}.

While the RAM focuses on data exchange via data connectors, so-called ``Value Adding Apps'' are also part of the specification~(see chapter 3 in~\citep{otto2023ram4}), and have been conceptualized since the first research project as pointed out by~\citet{otto2019designing}. According to the RAM, the concept of apps spans data processing apps and data transformation apps. Both types have in common that they are employed either before or after an actual data transfer, highlighting the tight relation to exchanged data. Our service architecture differs from this in that it facilitates service provision that is not directly related to a data exchange but considers services as an individual asset class, broadening the spectrum of possible offerings of a connector.

Besides, the RAM provides specifications for the distribution and installation of apps into a connector, via a dataspace-official so-called \emph{App Store}. Such concepts are not part of our work, however they might be applicable for generic servies and services that are what the RAM calls transformation apps. It should however be mentioned that there is currently no broadly adopted working implementation of an IDS App Store available. The upcoming fifth version of the RAM is available as a draft and mentions the term ``service providers''~\citep{steinbuss2024ram5}. Whether this is meant in the sense in that we use the term ``service'' remains unclear at the time of writing.

In addition, the IDSA develops the specification of the \emph{Dataspace Protocol} (\emph{DSP},~\citep{idsa2024dsp}) for communication between dataspace connectors.

\subsection{Relation to Gaia-X}
The Gaia-X European Association for Data and Cloud is another dataspace initiative with a different focus than the IDSA. According to chapter 3 of the architecture document, Gaia-X is concerned with ``decentralized digital trust'' whereas the IDSA focuses on sovereign data exchange and the DSP~\citep{gaiax2025architecture}. While Gaia-X addresses services as a first-class citizen since the beginning and actually considers data offerings as a sub-form of service offerings, it is conceptually not focused on supporting the implementation of such services. Concretely, Gaia-X provides the foundation for discovery and trusted interaction with services but leaves the technical aspects of interacting with the service to the provider and the consumer, especially not employing a connector.

\subsection{Relation to the Eclipse Data Space Components}
The \emph{Eclipse Data Space Components}, with their most prominent component, the EDC Connector, are a framework developed by a joint initiative of applied research institutions and industry under governance of the Eclipse Foundation. This software stack is built based on the specifications of IDSA and Gaia-X, including especially the aforementioned Dataspace Protocol~\citep{edcYYYeclipssite}. The key features of the EDC are to offer data assets bundled with usage policies, the trusted setup of usage contracts for them and the asset transfer.

The EDC is constructed as a modular system where a connector is built from a customized set of components. By its architecture it is open for extension. Two architecture components are integral to dataspace connectors according to the newest iteration of the DSP~\citep{idsa2025dsp_new}: \emph{Control Plane} and \emph{Data Plane}. The control plane is mainly responsible for the management of offers, the creation of contract agreements, and the \emph{coordination} of data transfers. It can be interacted with via the connector's \emph{management API}. The actual data transfer is facilitated by the data plane. There are various implementations of data planes available. More information on them can be found in~\citep{edc2025adopters}. Separating control and data flow follows the well-established software design pattern \emph{separation of concerns}, attributed to~\citet{dijkstra1982separation}.

The default HTTP data plane implementation for interacting with REST endpoints offers a feature for the consumer to parameterize a data transfer from the provider~\citep{edc2024dphttp}. This can be used to realize some of the potential applications of our service architecture, e.g., services that offer parameterized access to an SQL database, albeit without the elaborate state tracking and adaptability of our service architecture. A rather recent development is the decision of the developer community to implement a back channel for the data consumer to send feedback to the provider, e.g., errors~\citep{edc2024biditransfer}. This feature bears the potential to be used as a backend for the proposed service layer but does not provide similarly easily adoptable opportunities as-is.

\subsection{Further Related Works}
Conceptually, our work is related to the idea of \emph{Remote Procedure Call (RPC)}, defined in~\citep{rfc1057,rfc5531}.
RPC follows a client-server model whereas we do not distinguish between a server and a client but rather propose an abstraction layer that can assume both, the service provider and consumer role. Besides, our architecture works fully asynchronously and is built on top of the dataspace-established trust between the involved parties. The mechanism for invocation of a remote function, based on its name and its arguments, is similar.

Earlier, so called \emph{web services} have been formalized by a W3C working group, developed for supporting ``interoperable machine-to-machine interaction over a network''~\citep{booth2004webservicearch}. This approach is similar to ours in that it also defines interfaces for automated remote service usage with clear communication protocols. Besides, it includes the concept of a service registry via \emph{Universal Description, Discovery and Integration (UDDI)}, comparable to the dataspace catalog in our case. Usage policies and contracts as well as the trust infrastructure of dataspaces, however, are missing in this approach.

At the time of writing, we are not aware of any similar service architecture proposed for dataspaces beyond what has been discussed here.

\section{Architecture Concept}
\label{sec:architecture}
In this section we describe the individual aspects of the service architecture, from the roles dataspace participants assume in it to the core components. We consider a service as a mapping $f$ that maps $n$ input arguments $x_{i}$ to an output value $y$, i.e., $f: \left(x_{1}, \ldots, x_{n}\right) \mapsto y$, where $y$ might be a list of returned values. The $x_{i}$ are provided by the service consumer and sent to the provider, which sends $y$ back to the consumer.

Our abstraction layer thus has two distinct roles: The service consumer and the service provider. Still, each dataspace participant can assume both roles at the same time for different services and the same abstraction layer is used for both sides. We propose to abstract any technicalities of the service execution and asynchronous communication between provider and consumer. No communication with the consumer needs to be handled by the service itself.

Before explaining the state management and the inter-connector communication, we first introduce the interfaces of our abstraction layer.

\subsection{Invocation and Execution Interface}
\label{subsec:arch_extinterface}
As we consider services as first-class citizen assets with associated usage policies, we define that a service invocation requires an existing usage contract for this service, referenced by an alpha-numeric \texttt{contractId} known to both parties of the contract. We postpone further details on contracting to the EDC-specific implementation section, see \cref{subsec:impl_assetcontract}.

On the consumer side, the connector needs to provide three interface functions: The first starts a service invocation based on a \texttt{contractId} and the service arguments and returns an \texttt{invocationId} for later reference. The second allows the connector's user or interacting software to determine the state of an invocation based on its \texttt{invocationId}. Possible return values are the states defined later in this section. The third function consumes the \texttt{invocationId} as well and returns the result of the service invocation once it has finished, or a corresponding error message.

On the provider side, services are run directly within the connector or can be implemented as external code called by a wrapper service. To offer a service, providers need to implement the following three functions:

First, a service has to provide metadata about itself, consisting of
\begin{itemize}
    \item the (at least connector-)unique service id,
    \item an entry point for the service abstraction layer to start the execution,
    \item a list of the argument types of the service,
    \item and the service's return type.
\end{itemize}
The latter two properties could be replaced by only the number of arguments, however they may prove useful in optimized implementations, see \cref{subsec:impl_assetcontract}.

The second function receives the invocation arguments and is responsible for parsing them and possibly returns an error. We remark that the service implementation is responsible for processing the data safely and a safe execution. Best practices like input sanitization should be employed where applicable.

Finally, the third function is supposed to provide the main code of the service, equipped with local variables from the previous step. It shall return the service's result or indicate failure.

In \cref{sec:evaluation}, we demonstrate the usage of the consumer and provider interface with an example.

\subsection{State Management and Communication Between Connectors}
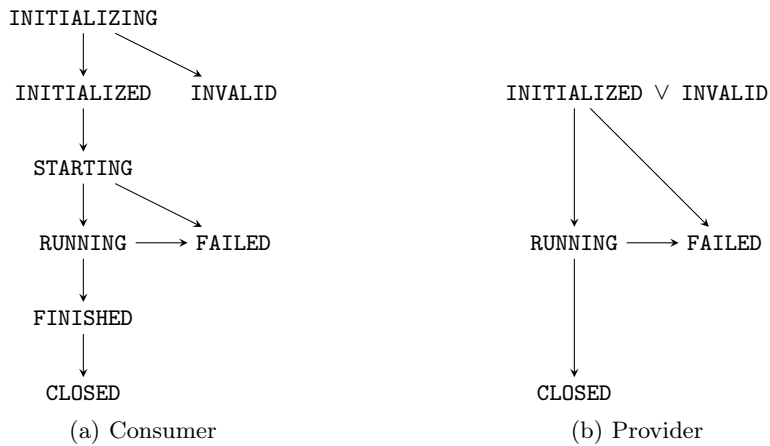
\begin{figure}
    \centering
    \subfigure[Consumer]{
        \begin{tikzpicture}[>=stealth]
            \node (initializing) at (0,0)   {\texttt{INITIALIZING}};
            \node (initialized)  at (0,-1)  {\texttt{INITIALIZED}};
            \node (invalid)      at (2, -1) {\texttt{INVALID}};
            \node (starting)     at (0, -2) {\texttt{STARTING}};
            \node (running)      at (0, -3) {\texttt{RUNNING}};
            \node (failed)       at (2, -3) {\texttt{FAILED}};
            \node (finished)     at (0, -4) {\texttt{FINISHED}};
            \node (closed)       at (0, -5) {\texttt{CLOSED}};
            
            \draw[->] (initializing) -- (initialized);
            \draw[->] (initializing) -- (invalid);
            \draw[->] (initialized) -- (starting);
            \draw[->] (starting) -- (running);
            \draw[->] (starting) -- (failed);
            \draw[->] (running) -- (finished);
            \draw[->] (running) -- (failed);
            \draw[->] (finished) -- (closed);
        \end{tikzpicture}
    }\qquad\qquad\qquad\qquad%
    \subfigure[Provider]{
        \begin{tikzpicture}[>=stealth]
            \node (initialized)  at (0,-1)  {\texttt{INITIALIZED}};
            \node (invalid)      at (2, -1) {\texttt{INVALID}};
            \node (running)      at (0, -3) {\texttt{RUNNING}};
            \node (failed)       at (2, -3) {\texttt{FAILED}};
            \node (closed)       at (0, -5) {\texttt{CLOSED}};
            
            \path (initialized) -- node[midway]{$\vee$} (invalid);
            \draw[->] (initialized) -- (running);
            \draw[->] (initialized) -- (failed);
            \draw[->] (running) -- (failed);
            \draw[->] (running) -- (closed);  %
        \end{tikzpicture}
    }
    \caption{State transition diagram for consumer and provider.}
    \label{fig:statetransitions}
\end{figure}

Internally, the asynchronous, remote service execution is organized by a decentralized state tracking of the invocation. The communication between the connectors is based on signals as opposed to a request-response scheme. This simplifies the overall logic and avoids the need to wait for responses. Also, no polling for status updates is required, benefiting the time from invocation to result as well as reducing network traffic. This design is in line with the architecture of the DSP~\citep{idsa2024dsp}. One aspect to consider regarding the signaling-based approach is that, in networks, timing anomalies may affect the arrival time of messages. We mitigate this by discarding any incoming backward transitions through the state graph depicted in \cref{fig:statetransitions}.

Consequently, both the provider and the consumer store their current information about the state of each service invocation. State transitions occur on three occasions: a) the consumer side invokes a service or retrieves its result, b) a signal is received from the other involved connector or c) on the provider side, the argument parsing or service execution triggers a state change.

\begin{figure}[p!]
    \centering
\begin{tikzpicture}[>=stealth]
    \coordinate (ctop) at (0,0);
    \coordinate (cbottom) at (0,8);
    \coordinate (ptop) at (5.5,0);
    \coordinate (pbottom) at (5.5,8);
    \draw[orange] (ctop) -- (cbottom) node[above]{\textbf{Consumer}};
    \draw[orange] (ptop) -- (pbottom) node[above]{\textbf{Provider}};
    
    \draw[->,blue] ($(cbottom)!.15!(ctop)$) -- node[above,midway]{ServiceInvocationSignal} ($(pbottom)!.15!(ptop)$);
    \draw ($(cbottom)!.15!(ctop)$) node[left]{\texttt{INITIALIZING}};
    \draw ($(pbottom)!.15!(ptop)$) node[right,align=left]{\texttt{INITIALIZED}\\/ \texttt{INVALID}};
    
    \draw[<-,darkgreen,dashed] ($(cbottom)!.225!(ctop)$) -- node[above,midway]{ServiceInitializedSignal} ($(pbottom)!.225!(ptop)$);
    \draw[<-,darkred,dotted] ($(cbottom)!.3!(ctop)$) -- node[above,midway]{ServiceInvalidSignal} ($(pbottom)!.3!(ptop)$);
    \draw ($(cbottom)!.225!(ctop)$) node[left]{\texttt{INITIALIZED}};
    \draw ($(cbottom)!.3!(ctop)$) node[left]{\texttt{INVALID}};
    
    \draw[->,blue] ($(cbottom)!.5!(ctop)$) -- node[above,midway]{ServiceExecutionSignal} ($(pbottom)!.5!(ptop)$);
    \draw ($(cbottom)!.5!(ctop)$) node[left]{\texttt{STARTING}};
    \draw ($(pbottom)!.5!(ptop)$) node[right,align=left]{\texttt{RUNNING}\\/ \texttt{FAILED}};
    
    \draw[<-,darkgreen,dashed] ($(cbottom)!.575!(ctop)$) -- node[above,midway]{ServiceRunningSignal} ($(pbottom)!.575!(ptop)$);
    \draw[<-,darkred,dotted] ($(cbottom)!.65!(ctop)$) -- node[above,midway]{ServiceFailedSignal} ($(pbottom)!.65!(ptop)$);
    \draw ($(cbottom)!.575!(ctop)$) node[left]{\texttt{RUNNING}};
    \draw ($(cbottom)!.65!(ctop)$) node[left]{\texttt{FAILED}};
    
    \draw[<-,blue] ($(cbottom)!.85!(ctop)$) -- node[above,midway]{ServiceFinishedSignal} ($(pbottom)!.85!(ptop)$);
    \draw ($(cbottom)!.85!(ctop)$) node[left] (cfinished) {\texttt{FINISHED}};
    \draw ($(pbottom)!.85!(ptop)$) node[right]{\texttt{CLOSED}};
    
    \draw ($(cbottom)!.925!(ctop)$) node[left] (cclosed) {\texttt{CLOSED}};
    \draw[->] (cfinished.west) to[bend right=42] (cclosed.west);
    
    \end{tikzpicture}
     \caption{State sequence diagram for the service execution for consumer and provider. The transition from \texttt{FINISHED} to \texttt{CLOSED} on the consumer side is triggered once the result has been retrieved by the participant operating the connector. Standard signals are solid blue. In cases where there can be a positive or a negative feedback, they are green dashed or red dotted, respectively.}
    \label{fig:states}
\end{figure}
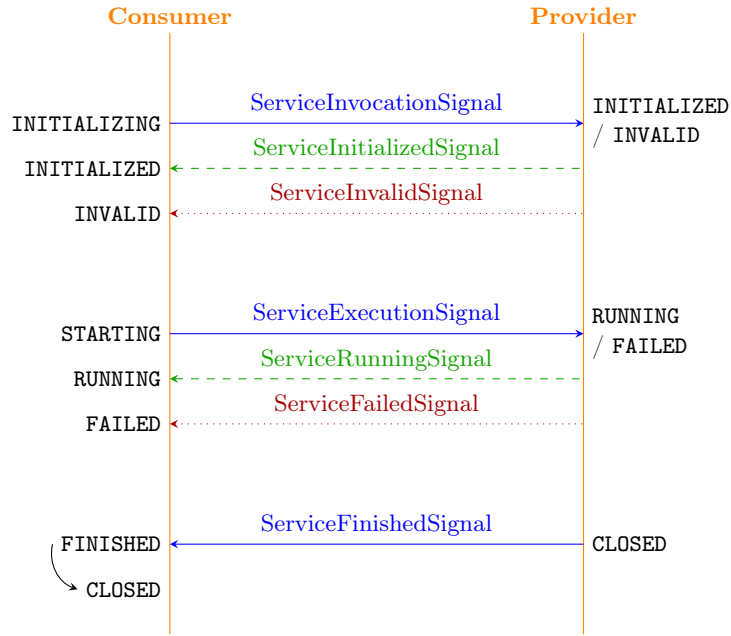

We define the states as presented in \cref{tab:states}. Two figures help understanding the following description of the service execution state management: the relation of states and signaling can be seen in \cref{fig:states}. \Cref{fig:statetransitions} provides a reduced view on the states only and the possible transitions, separated for the consumer and the provider side.

\begin{table}[p]
    \caption{Service Invocation States. C is the consumer, P the provider.}
    \label{tab:states}
    \centering
    \begin{tabularx}{\textwidth}{lXc}
        \toprule
        \textbf{State} & \textbf{Description} & \textbf{Role} \\\midrule
        \texttt{INITIALIZING} & initial state, awaiting initialization result from provider & C\\\midrule
        \texttt{INITIALIZED} & invoked service exists at the provider's connector and is associated with a valid contract & P/C \\\midrule
        \texttt{INVALID} & service does not exist or the \texttt{contractId} is invalid & P/C \\\midrule
        \texttt{STARTING} & service execution start signaled to provider, awaiting feedback & C \\ \midrule
        \texttt{RUNNING} & arguments parsed successfully, execution started & P/C \\\midrule
        \texttt{FAILED} & failure during argument parsing or during execution with an exception & P/C \\\midrule
        \texttt{FINISHED} & service execution finished (successfully or with error) & C \\\midrule
        \texttt{CLOSED} & result sent to the consumer (P) / result retrieved via API (C) & P/C \\\bottomrule
    \end{tabularx}
\end{table}

Upon invocation of some service, a new \texttt{invocationId} is assigned and stored with the state \texttt{INITIALIZING}. The provider is sent a \texttt{ServiceInvocationSignal}, containing the \texttt{invocationId} and the associated \texttt{contractId}. On the provider side, the \texttt{contractId} is validated and it is verified that the corresponding service is properly loaded. If both is the case, the \texttt{invocationId} is stored with the state \texttt{INITIALIZED} and a \texttt{Service\-InitializedSignal} is sent to the consumer. If not, the invocation is set to the \texttt{INVALID} state and an according \texttt{ServiceInvalidSignal} is sent, containing an error message. %

When the invocation transitions to the \texttt{INITIALIZED} state on the consumer side after receiving the corresponding signal, the connector automatically sends a \texttt{ServiceExecutionSignal} containing the arguments for the service's execution, cached from when the invocation has been created, and advances it to the \texttt{STARTING} state.

Upon receipt of the \texttt{ServiceExecutionSignal}, the provider verifies the number of arguments and then forwards them to the aforementioned service-specific function for parsing and saving them. If an error occurs during this process, the invocation moves to the \texttt{FAILED} state and a \texttt{ServiceFailedSignal} is sent to the consumer, containing an error message.

In case the prior step has succeeded, the invocation is moved to the \texttt{RUNNING} state on the provider side, which is signaled via a \texttt{ServiceRunningSignal} to the consumer. The actual execution is non-blocking and performed by invoking the service's execution method implementation.

In case the execution fails, the behavior is the same as for an error during the execution preparation procedure: the invocation is set to \texttt{FAILED} and the consumer is notified accordingly, accompanied by an error message. To reduce complexity, we do not define distinct states for failures during the argument parsing and execution. Thus, the error message should be sufficiently expressive to distinguish both. Given no fatal error has occurred during execution, the service execution eventually exits normally. Then, the provider passes the information and the actual result to the consumer via a \texttt{ServiceFinishedSignal}. This procedure is also followed if the service has finished with an error and no data to return as long as there is no uncaught exception. As this represents the last step for the provider, the invocation is moved to the \texttt{CLOSED} state.

On the consumer side, the service's result is cached. Therefore, the invocation first assumes the \texttt{FINISHED} state and does not advance to \texttt{CLOSED} immediately as done at the provider. After that, the participant may invoke the connector's above mentioned function for retrieving the service result, i.e., the resulting data or the error message. The connector retrieves the result from its cache, removing it, and moves the invocation to the terminal \texttt{CLOSED} state. Both sides may decide to keep the final state information for documentation purposes.

\section{Implementation for the EDC}
\label{sec:impl}
In this section we describe relevant aspects of our prototypic implementation of the service architecture for the EDC. Technically, the abstraction layer is realized as an extension building on the EDC's extension interface. We structure this section corresponding to the previous one, complemented by an EDC-specific part on contracting.

\subsection{Invocation And Execution Interface}
\label{subsec:impl_extinterface}
On the consumer side, we implement the three required functions defined in \cref{subsec:arch_extinterface} by extending the EDC's management API with additional endpoints under the prefix \texttt{/serviceinvocation}: the \texttt{/invoke} POST endpoint realizes the invocation function and is thus parameterized with the \texttt{contractId}, returning the \texttt{invocationId}, which is implemented as a random UUID~\citep{rfc9562}. It consumes the service arguments as a JSON-encoded array. The \texttt{/status} GET endpoint implements the status retrieval function, based on the \texttt{invocationId}. For implementing the third function, the result retrieval, we provide the \texttt{/result} GET endpoint, which is by definition parameterized with the \texttt{invocationId} as well. In case of success, our implementation does not only return the result but its type as well. To facilitate polling-free setups, the \texttt{/invoke} endpoint has an optional parameter for submitting a URL that is called once the service is finished. This URL might point to an endpoint, which then triggers the result transfer.

On the provider side, services are expected to be provided as extensions to the EDC, written in a \textit{Java}-interoperable language such as \textit{Kotlin} or \textit{Java}. When the connector is built, service extensions are compiled into the binary. For more flexibility, the wrapper for external services described before can be implemented as a generic service that invokes external code, e.g., via execution of system processes for locally executed languages or via HTTP requests.

\begin{figure}[ht]
    \centering
    \begin{tikzpicture}
        \begin{interface}[text width=6.6cm]{Service}{0,0}
            \operation{getMetadata(): ServiceMetadata}
            \operation{loadArgs(args: Array<String>?): Unit}
            \operation{execute(): ServiceResult}
        \end{interface}
    \end{tikzpicture}
    \caption{The methods of the \texttt{Service} interface that services must implement. \texttt{Unit} in Kotlin corresponds to \texttt{void} in Java.}
    \label{fig:serviceinterface}
\end{figure}
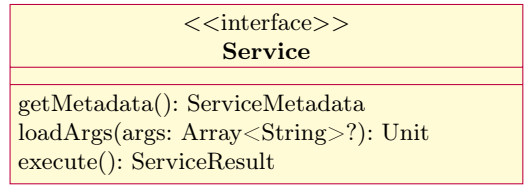

The three methods defined above that a service needs to offer, have to be provided by implementing the interface depicted in \cref{fig:serviceinterface}, with the \texttt{getMetadata} method implemented as a static class method. This method must return an instance of \texttt{ServiceMetadata} with the following implementation-specific aspects:
\begin{itemize}
    \item We define the service's id as the service extension's main package.
    \item The entry point is referenced by the service's main class that needs to be instantiated for execution.
\end{itemize}

The \texttt{loadArgs} method receives the invocation arguments. Instead of passing them as directly as a JSON array, we pre-parse them into a list of strings, each containing one JSON-encoded argument of the service. In our implementation, errors that occur while this method parses the arguments into local fields have to be thrown as exceptions. Service implementers should consider that the exception message is sent to the consumer as error message. Finally, the \texttt{execute} method implementation is the one with the actual service execution code. It returns the result as a \texttt{ServiceResult} consisting of a flag indicating invocation success and -- in case of success -- the returned data. We do not enforce some return format but propose to use JSON.

\subsection{State Management and Communication Between Connectors}
To implement the decentralized state tracking, we augment the EDC's relational database with an additional table mapping \texttt{invocationId}s to the current state and additional metadata like the direction of the invocation, i.e., incoming or outgoing, or the associated \texttt{contractId}.

A key-value map is used to cache the service arguments in-memory between the \texttt{/invoke} management API call and the actual passing of arguments to the service provider. Regarding the non-blocking service execution, we opted for an implementation based on an $n$-threaded executor pool, equipped with callbacks for passing back the result or a potential error.

As for the arguments, the service result caching on the consumer connector is done in-memory in our implementation. In future implementations of the abstraction layer, a persistent result cache can be added, allowing to retrieve the result even after a restart of the connector. Besides, upon connector start all incoming service invocations, i.e., invocations where the connector assumes the provider role, that are not yet \texttt{CLOSED} are moved to \texttt{INVALID} state as they cannot be processed further as arguments as well as execution state are lost. The consumer is notified about this state change as well. In contrast, a reboot of the consumer might be acceptable, e.g., for long running services.

\subsection{Asset and Contract Setup}
\label{subsec:impl_assetcontract}
As outlined in \cref{sec:relatedwork}, data exchange in EDC-based dataspaces requires a contract between the two involved parties. The structure for creating data offers is as follows: the provider creates a so-called asset that consists of metadata to appear in the dataspace's catalog and information for the connector on how to access the asset for later transfer. The latter can, e.g., be given as a REST API endpoint in the provider's infrastructure and is not shared with the data consumer. The provider might define further private properties not shared with other participants. All properties need \emph{Uniform Resource Identifier}s (\emph{URI}s) as keys. An offer to other dataspace participants is formed by coupling an asset with an access policy and a usage policy, both specified in the \emph{Open Digital Rights Language (ODRL\emph{, see~\citep{iannella2018odrl})}}.

We aim at integrating with this infrastructure as seamlessly as possible to ease the adoption of our service layer as well as to benefit from the existing tooling for well-authenticated, regulated access. Thus, the following procedure is required to provide a service: first, as described in \cref{subsec:impl_extinterface}, the extension for the service needs to be compiled into the binary. Once the connector is running, an asset has to be created, describing the service. For the service to be usable, the number of arguments it consumes and their types need to be specified in the asset's properties. We propose to do this in form of a JSON array, with \emph{MIME types}, allowing for future machine processing. The result type is also of interest to the consumer, albeit also shareable with the service result later. In our proof-of-concept implementation, we defined custom URIs for this -- in the longer run, standardization is favorable. Apart from these properties, no mandatory public properties are required. However, we require a private property that indicates the \texttt{serviceId} introduced in \cref{subsec:arch_extinterface}, allowing to associate the asset and derived contracts with the service implementation. A definition for how to access associated data is not required for such assets by construction. In later versions of our implementation, the argument types as well as the return type may be automatically added to the asset metadata based on the \texttt{serviceId} and the metadata directly provided by the service extension.

In the context of asset creation, we introduced an additional property to indicate the asset type, for clearly separating classical data assets and services. However, this is not strictly needed.

Finally, an access policy, defining which connectors can see the offer, and a usage policy, i.e., how the service may be used, need to be defined. Bundled as a contract offer, the service is prepared to be used by consumers. Consumers need to negotiate a usage contract with the provider. They can then use the resulting \texttt{contractId} for invoking the service as described in \cref{subsec:arch_extinterface}.

\section{Evaluation}
\label{sec:evaluation}
We experimentally evaluate our service architecture by implementing a service based on our proof-of-concept implementation of the abstraction layer and analyze the runtime behavior, demonstrating the practical applicability.

\subsection{MNIST Image Recognition Service}
\label{subsec:evalmnist}
Specifically, we implemented an image recognition service. For simplicity, we opted for training a simple \textit{Convolutional Neural Network (CNN)} on the well-known MNIST dataset~\citep{lecun1998mnist} and developing a service for detecting digits drawn in an HTML \textit{Canvas}~\citep{whatwg2025canvas}.

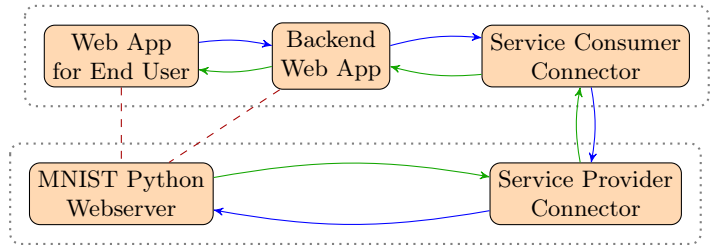
\begin{figure}[ht]
    \centering
    \tikzstyle{component} = [rectangle, draw=black, rounded corners, text centered, fill=orange!30, align=center]
    \tikzstyle{bentarrow} = [bend left=10]
    \begin{tikzpicture}[node distance=2cm,>=stealth']
        \node (webapp) [component] {Web App\\for End User};
        \node (backend) [component,right of=webapp,anchor=west] {Backend\\Web App};
        \node (ccon) [component,right of=backend,anchor=west] {Service Consumer\\Connector};
        \node (pcon) [component,below of=ccon,anchor=south,yshift=-.25cm] {Service Provider\\Connector};
        \node (mnist) [component,below of=webapp,anchor=south,yshift=-.25cm] {MNIST Python\\Webserver};
        
        \draw[dotted,thick,black!50, rounded corners] ($(webapp.north west)+(-.25,.25)$) rectangle ($(ccon.south east)+(.25,-.25)$);
        \draw[dotted,thick,black!50, rounded corners] ($(mnist.north west)+(-.25,.25)$) rectangle ($(pcon.south east)+(.25,-.25)$);
        
        \draw[->,blue] (webapp) to[bentarrow] (backend);
        \draw[->,darkgreen] (backend) to[bentarrow] (webapp);
        
        \draw[->,blue] (backend) to[bentarrow] (ccon);
        \draw[->,darkgreen] (ccon) to[bentarrow] (backend);
        
        \draw[->,blue] (ccon) to[bentarrow] (pcon);
        \draw[->,darkgreen] (pcon) to[bentarrow] (ccon);
        
        \draw[->,blue] (pcon) to[bentarrow] (mnist);
        \draw[->,darkgreen] (mnist) to[bentarrow] (pcon);
        
        \draw[dashed,darkred] (webapp) to (mnist);
        \draw[dashed,darkred] (backend) to (mnist);
    \end{tikzpicture}
    \caption{Data Flow in the MNIST experiment setup: blue indicates the image passed through the subsequent layers, green indicates the resulting classification being passed back to the Web App. The dashed red lines indicate that there is no direct communication between the end user application or its backend and the MNIST service. The gray dotted rectangles indicate the domains of service provider and consumer.}
    \label{fig:mnistexample}
\end{figure}

In detail, we implemented a web app that presents the user with a canvas for drawing a digit and displays the result after recognition. The detailed process is also depicted in~\cref{fig:mnistexample}. A button starts the service invocation by sending the drawing to the backend. From there, the invocation endpoint of the EDC management API is called as explained in \cref{subsec:impl_extinterface}. Subsequently, the encoded digit image is forwarded to the provider's connector that uses it to invoke the recognition model, which is served via a Python-based web server. This means that the execution method of our service calls the recognition API and returns once this request is concluded via a result. The web app polls the consumer connector for the state of the service. Once the state arrives at \texttt{FINISHED}, the result is retrieved as described in \cref{subsec:impl_extinterface} and displayed to the user. In case of an error, the error state and a corresponding message is shown.

We remark that the associated data flow as depicted in \cref{fig:mnistexample} does not involve any direct connections between the technical service consumer, i.e., the web app, and the actual service. This improves the findability of the service compared to a direct service offering and allows the model operator to restrict the network access for the service to its own EDC Connector.

The MNIST service apparently is a rather simple service with low practical relevance. However, it can be trivially adapted to more elaborate scenarios, like a mobile application for classifying photos in medical contexts. The backend of such an application, attached to a connector, can invoke a service offered by a company that, e.g., specializes on skin wound classification and offers a related service via the dataspace. This use case benefits from the usage policies, the technically implemented trust and the dataspace governance structure. In a conventional dataspace setup, the classification model developer would need to share the model with the application provider, which might be incompatible with the developer's business interests as well as legal regulations.

\subsection{Runtime Compared to Direct Service Invocation}
For evaluating the practical implications of our service architecture, we evaluate the latency of our method compared to a direct invocation of the service. Our setup for this is as follows: as the service being invoked, we use the MNIST recognition service introduced before. The model has roughly 1.2M parameters and is executed on a server CPU due to organizational constraints. While this is suboptimal compared to running on a GPU, it is bearable for a small model like the one used. For direct invocation of the service, we use an HTTP request to the Python-based server that evaluates the model, sent by the webapp's backend.

For the connector-based invocation, we slightly modify the system from \cref{subsec:evalmnist}. Concretely, we make use of the feature for passing an endpoint at invocation time that is called once the service finishes. When this endpoint is contacted by the consumer connector, we immediately fetch the result in the web app's backend with an additional request to the connector, bypassing the explicit result retrieval triggered by the frontend. This prevents additional delay introduced by polling from the frontend or additional complexity for establishing a mechanism for notifying the frontend. In a real use case, one would in most cases opt for the latter. We measure the time between calling the connector's \texttt{/invoke} endpoint and successful retrieval of the result via the \texttt{/result} endpoint. The frontend is displayed on a local end-user machine. The consumer connector and the MNIST client, i.e., the webapp and its backend, are running on a hosted virtual machine (VM) while the provider connector alongside the MNIST service server are running on another hosted VM, operated by a different provider to evaluate under realistic conditions. All components are \emph{dockerized}. This setup closely resembles realistic setups with the exception that the client and the actual services typically would run on separate machines on the dataspace participants' own infrastructure.

For direct and service architecture-based invocation, we performed ten evaluation runs with a ``hot setup'', i.e., after already calling the service once after startup. The timings are depicted in \cref{tab:timing}. We measure from the consumer webapp backend instead of the frontend to eliminate differences due the local machine's internet connection. This affects the timings for both invocation forms identically. Direct invocation is faster than EDC-mediated invocation by a margin. Concretely, invoking the service via our service layer is slower by a factor of $10.64$. A significant performance impact, however, has to be expected due to the additional networking interactions and the additionally processed code including database accesses for state management. \Cref{tab:timing} further shows that the largest part except for about \millisec{23} is consumed by the main service invocation part. The result retrieval is performed in the remaining time. With the used hardware, the actual neural network inference consumes \millisec{7.7} on average.

\begin{table}[ht]
    \caption{Time Consumption for EDC-mediated Service Execution}
    \label{tab:timing}
    \centering
    \begin{tabularx}{\textwidth}{Xrr}
        \toprule
        \textbf{Time Span} & \textbf{Mean Duration} & \textbf{Standard Deviation $\sigma$} \\\midrule
        start until \texttt{FINISHED} & \millisec{245.0} & \millisec{66.46} \\\midrule
        \texttt{FINISHED} until result requested & \millisec{14.7} & \millisec{6.72} \\\midrule
        result requested until retrieved & \millisec{8.5} & \millisec{1.35} \\\midrule
        \textbf{complete: start until result retrieved} & \millisec{268.2} & \millisec{68.47} \\\midrule\midrule
        direct invocation & \millisec{25.2} & \millisec{13.43} \\\midrule\midrule
        neural network inference only & \millisec{7.7} & \millisec{5.45} \\\bottomrule
    \end{tabularx}
\end{table}

Although significantly slower than direct invocation, invoking the MNIST service via the EDC still consumes only roughly a quarter of a second. The relative runtime difference decreases with complexity of the underlying service. While higher for simple format transformations, it becomes negligible for long-running services like, e.g., complex SQL operations or complicated LLM-based services.

\section{Discussion}
\label{sec:discussion}
The proposed service architecture for dataspaces has proven its feasibility by means of a proof-of-concept implementation for the EDC alongside the successful demonstration with an example in \cref{sec:evaluation}. Still, there are aspects that justify critical discussion.

On the implementation side, we made use of the \emph{sovity Messenger} as introduced by the company \emph{sovity} in their \emph{EDC Community Edition}\footnote{More information can be found at \url{https://github.com/sovity}.}.
While this allowed for a fast implementation of our abstraction layer, newer versions of this library are under the more restrictive \emph{Elastic license}~\citep{elasticYYYYlicense}, compared to the previously used \emph{Apache license}~\citep{apache2004license}, potentially preventing broader adoption. As there is the widely employed DSP for the communication between dataspace connectors, standardized by the IDSA and implemented by the EDC, the ultimate goal must be to also standardize the service-related communication and integrate it into the DSP, after discussion and potential refinement within the scientific and application community.

As of now, offering a service requires creating an EDC extension and a corresponding contract offer for the catalog, with properties to be set via multiple API requests. The prior might present a significant entry barrier that warrants further research. For instance, one could develop a generic base service for format transformations, which can then be customized using low-code and offered without explicit extension development by the provider. Similar approaches would lower the barrier for established database systems. Regarding the creation of the offers with their properties, missing frontend support currently limits the ease of adoption.

On the consuming side, supportive tooling might in practice prove even more important than on the provider side. After setting up the usage contract, the connector API interactions explained in \cref{subsec:impl_extinterface} need to be performed by the consumer. When integrating such a service usage directly into an application of the consumer, this is sufficient. However, there might be use cases where an end user needs to use the service manually. For this, a generic frontend that allows for passing arguments and invoking the service lowers the adoption barrier.

Another topic to be discussed when it comes to remote service invocation is security. As pointed out earlier, the safe execution of a service is a responsibility of its provider. Input sanitization and validation are crucial aspects in this regard and can mitigate threats like SQL injection. At the abstraction layer level, however, there exist measures as well to support safe execution: our solution validates the number of arguments passed to the service. It does not validate their type to maintain flexibility for service providers. Besides, the separation of argument loading and service execution as introduced in \cref{subsec:arch_extinterface} reduces the attack surface by design. Execution in a separate thread is an implementation-specific measure aiming at the same goal. It can be improved upon by further encapsulation in future, hardened implementations of the execution engine.

In general, we see a broad scope for adoption with various kinds of services. The spectrum reaches from ``improved file transfers'' to complex services: our service abstraction layer can be used to provide access to multiple underlying files via one contract, with a service that returns data based on some identifier. A use case for this could be large image datasets. Another category of services are parameterized query interfaces for databases. Beyond this, one can also build services with complex inner logic like executing machine learning models as presented in \cref{subsec:evalmnist} or running forecast models given input parameters. In summary, this allows dataspaces to also span the service layers of the tree architecture introduced by~\citet{gillmann2024dsanalytics}.

Besides, our architecture can be transferred to dataspaces employing other connectors than the EDC via adapted implementations.

\section{Conclusion}
\label{sec:conclusion}
We have introduced a novel abstraction layer for providing services in dataspaces, implemented using the well established EDC connector. The proposed architecture lays the conceptual and technological foundation for offering arbitrary, parameterized services while exploiting the benefits of dataspaces, including technically established trust and an already in place architecture for usage policies and contracts. By allowing to produce results based on internal data without the need to share them with service consumers, this method takes the data sovereignty principle of dataspaces even further. We also have demonstrated the practical applicability of our architecture by implementing and evaluating an image recognition service with a performance hit of about \SI[parse-numbers=false]{\frac{1}{4}}{\second} compared to plain execution of the service without dataspace integration.

Based on the proposed abstraction layer, we see a broad field of possible applications and future research, including standardization of the inter-connector communication for service execution and development of generic base services. A field for improvement is the reduction of latency introduced by the connector, e.g., by optimizing database access.

In addition to the aspects discussed in \cref{sec:discussion}, also an in-depth comparison from the perspective of web services as defined by~\citet{booth2004webservicearch} and later work warrants further research given our work mostly focuses on the perspective of dataspaces.

\printbibliography
\end{document}